\documentclass[a4paper,twocolumn,notitlepage,nofootinbib,longbibliography,superscriptaddress,floatfix]{revtex4-2}
\usepackage{
    adjustbox,
    algorithm,
    algpseudocode,
    amsmath,
    amssymb,
    amsthm,
    booktabs,
    braket,
    comment,
    csquotes,
    enumitem,
    graphicx,
    IEEEtrantools,
    mathtools,
    multirow,
    natbib,
    physics,
    refcount,
    relsize,
    url,
    times
}

\usepackage[english]{babel}
\usepackage[normalem]{ulem}
\usepackage[caption=false]{subfig}
\usepackage{algorithm}
\usepackage{algpseudocode}

\usepackage[dvipsnames]{xcolor}
\usepackage[colorlinks=true, urlcolor=Maroon, linkcolor=Emerald, citecolor=Periwinkle]{hyperref}

\providecommand{\to}{\ensuremath{\Tilde{o}}}

\providecommand{\calD}{\ensuremath{\mathcal{D}}}

\providecommand{\calY}{\ensuremath{\mathcal{Y}}}

\providecommand{\bbR}{\ensuremath{\mathbb{R}}}

\newtheorem{theorem}{Theorem}

\newtheorem{proposition}[theorem]{Proposition}

\newenvironment{remark}[1][Remark]{\begin{trivlist}
\item[\hskip \labelsep {\bfseries #1}]}{\end{trivlist}}

\newtheorem{innercustomgeneric}{\customgenericname}
\providecommand{\customgenericname}{}
\newcommand{\newcustomtheorem}[2]{%
  \newenvironment{#1}[1]
  {%
   \renewcommand\customgenericname{#2}%
   \renewcommand\theinnercustomgeneric{##1}%
   \innercustomgeneric
  }
  {\endinnercustomgeneric}
}

\newcustomtheorem{customtheorem}{Theorem}
\newcustomtheorem{customlemma}{Lemma}
\newcustomtheorem{customproposition}{Proposition}

\newcommand{\leiden}{Leiden University, Niels Bohrweg 1, 2333 CA Leiden, Netherlands}
\newcommand{\vw}{Volkswagen Group Innovation, Berliner Ring 2, 38440 Wolfsburg, Germany}

\begin{document}

\title{Uncertainty-aware quantum machine learning based on Dempster-Shafer theory}

\author{Marie Kempkes}
\email{marie.kempkes@volkswagen.de}
\affiliation{\leiden}
\affiliation{\vw}
\author{Jakob Spiegelberg}
\affiliation{\vw}
\author{Evert van Nieuwenburg}
\affiliation{\leiden}
\author{Vedran Dunjko}
\affiliation{\leiden}

\begin{abstract}
    A large body of work in quantum machine learning has centered on characterizing the expressivity and performance of quantum models.
    In contrast, little attention has been paid to the reliability of predictions beyond average performance measures such as accuracy.
    In this work, we introduce an uncertainty-aware quantum classification framework based on Dempster-Shafer theory.
    This theory allows to express uncertainty through weights assigned not only to individual classification labels but to sets of labels.
    We introduce a training strategy that enables parameterized quantum circuits to learn such assignments directly from data.
    This provides an efficient way to prepare the corresponding quantum states and enables the application of different post-processing strategies.
    In particular, it offers a practical setting for quantum Dempster-Shafer fusion procedures with polynomial resource requirements, even when the corresponding mass functions are defined over exponentially many subsets.
    Overall, our work represents a step toward enabling quantum models to express uncertainty about their predictions, thereby extending the evaluation of quantum models beyond predictive performance alone.
\end{abstract}

\maketitle

\section{Introduction} \label{s:introduction}   
    Quantum machine learning (QML) has emerged as a promising paradigm that leverages quantum mechanical principles to learn from data~\cite{biamonte2017quantum, dunjko2018machine, carleo2019machine, schuld2021machine, cerezo2021variational, bharti2022noisy}.
    Despite well-known challenges, including issues of trainability such as barren plateaus~\cite{mcclean2018barren} and the classical simulability of certain quantum models~\cite{cerezo2025does}, our understanding of quantum machine learning has improved significantly in recent years~\cite{anschuetz2025unified, gil2024understanding, gil2025relation}.
    A growing body of work has identified learning tasks for which QML offers provable advantages over classical approaches~\cite{liu2021rigorous, sweke2021quantum, huang2021information, gyurik2023exponential, pirnay2023superpolynomial, molteni2024exponential, barthe2025quantum}.

    However, as in classical machine learning, performance gains alone are insufficient to enable the deployment of quantum machine learning models in practice, especially, e.g., in high-risk applications. 
    Standard evaluation metrics such as accuracy capture average performance, but do not quantify the uncertainty associated with individual predictions.
    For example, in autonomous driving, a classifier that achieves high accuracy on average may still fail to recognize rare but critical situations, such as unusual obstacles or ambiguous sensor inputs.
    
    This limitation highlights the need for dedicated uncertainty quantification (UQ) methods.
    UQ aims to estimate a model’s uncertainty, i.e., what the model does \emph{not} know for a given input.
    A common distinction is between \emph{epistemic uncertainty}, which arises from incomplete knowledge of the underlying data-generating process and may be reduced through additional data or improved models, and \emph{aleatoric uncertainty}, which reflects ambiguity inherent in the observed data and may persist even for an optimal predictor~\cite{hullermeier2021aleatoric}.
    Access to such uncertainty estimates enables downstream safety mechanisms, e.g., abstention from high‑risk decisions if the prediction for a given input is considered highly uncertain.

    In classical machine learning, a variety of approaches to uncertainty quantification have been developed.
    Prominent examples include Bayesian neural networks~\cite{mckay1992apractical}, ensemble-based approaches~\cite{lakshminarayanan2017simpleandscalable}, and Monte Carlo dropout~\cite{gal2016dropout}.
    Methods based on Dempster-Shafer theory (DST), a generalized probability framework, offer a principled way to represent uncertainty through assignments to sets of outcomes~\cite{manchingal2022epistemic, manchingal2023random, kempkes2024reliable}.
    Interestingly, DST aligns well with the structure of quantum models, making it a strong candidate for uncertainty-aware QML~\cite{zhou2023bf}.
    
    \begin{figure}[t]\label{fig:summary_figure}
        \vspace{2.25\baselineskip}
        \includegraphics[width=1\columnwidth]{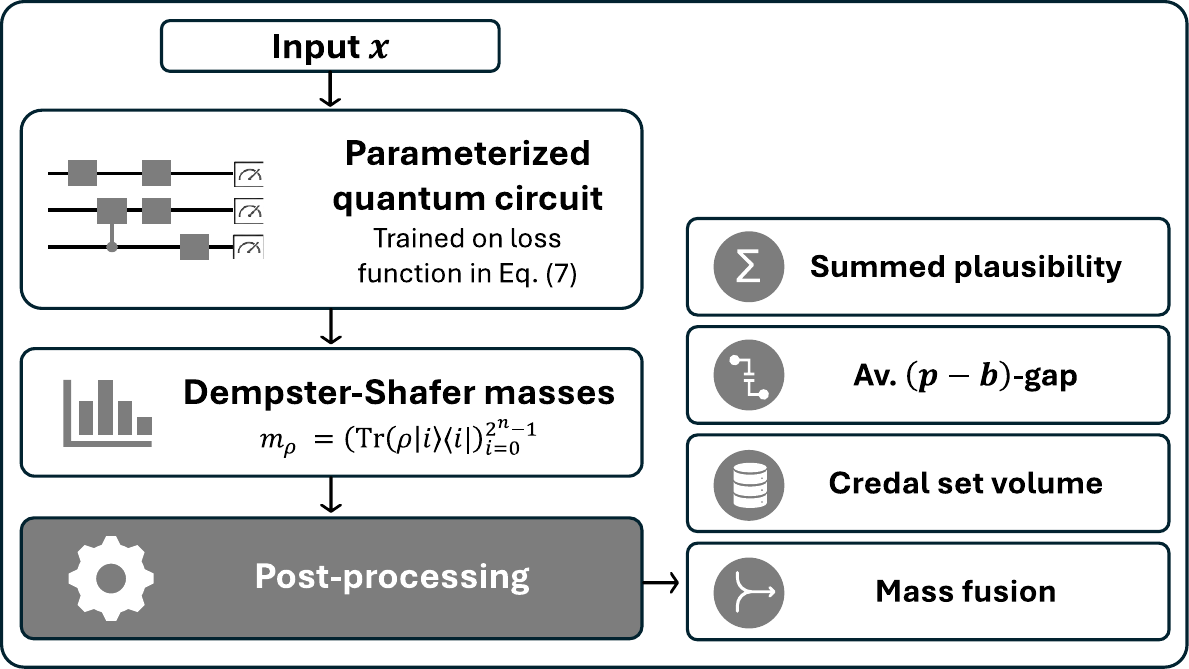}
    \caption{Conceptual overview of the proposed method. A parameterized quantum circuit is trained using the loss in Eq.~\eqref{eq:pl_sglloss}. The resulting basis state probabilities encode masses from Dempster–Shafer theory (see Section~\ref{s:trainingalgo}), which are subsequently used in different post-processing steps (Section~\ref{s:postprocessing}).}
    \end{figure}

    In this work, we introduce a framework that integrates Dempster-Shafer-based uncertainty quantification techniques into parameterized quantum circuits for classification tasks.
    A conceptual overview is shown in Fig.~\ref{fig:summary_figure}.
    We propose a classification model in which a parameterized quantum circuit is trained to produce generalized probability assignments, referred to as \emph{masses}, by encoding them into probabilities of basis states.
    We employ a training strategy from classical machine learning~\cite{kempkes2024reliable}, adapt the corresponding loss function to parameterized quantum circuits and show that its evaluation is efficient in the number of qubits.
    Additionally, we show that the resulting mass function, whose dimension grows exponentially with the number of qubits, is sparse in an idealized zero-loss setting and investigate numerically whether this behavior persists under more realistic training conditions.
    We further propose different post-processing procedures for efficient extraction of information about the uncertainty associated with individual predictions.
    We evaluate the resulting uncertainty measures numerically on the downstream task of out-of-distribution detection using real-world datasets.    
    Finally, we study settings in which multiple classifiers provide information about the same input.
    Dempster-Shafer theory provides dedicated rules for combining the resulting mass assignments, a procedure referred to as \emph{mass fusion}. 
    Recent work has shown that DST fusion operations can be implemented directly on quantum-encoded mass functions using polynomial quantum resources~\cite{luo2026attributefusion}, avoiding explicit manipulation of exponentially large mass vectors.
    Our framework complements these results by providing a training procedure through which such quantum representations arise naturally from the learning process.

\section{Preliminaries}\label{s:background}

    In this section, we review the foundations of Dempster-Shafer theory and its application to classification in classical machine learning.
    Related work is discussed in Appendix~\ref{s:relatedwork}.

\subsection{Dempster-Shafer theory}\label{s:DST}
    Dempster-Shafer theory (DST), also known as evidence theory, is a mathematical framework for reasoning under incomplete, imprecise, or contradictory information.
    Instead of assigning probabilities to individual outcomes, DST assigns generalized probability assignments, so-called \emph{masses}, to all subsets of the set of possible outcomes~\citep{dempster1967amultivalued}.
    This formulation allows uncertainty to be represented at the level of subsets and supports distinctions between uncertainty arising from lack of information and uncertainty arising from randomness, which are commonly associated with epistemic and aleatoric uncertainty, respectively.
    
    To illustrate this, consider tossing a coin that is known to be fair: both heads and tails have a probability of a half. 
    If nothing is known about the fairness of the coin, DST represents this state of ignorance by assigning a mass of $1$ to the set $\{{heads}, {tails}\}$ and a mass of $0$ to all other subsets.
    In this way, it refrains from making unjustified assumptions about the fairness of the coin.

    We now introduce the basic objects of Dempster–Shafer theory.
    Let $\calY = \{1,...,n\}$ denote a \emph{frame of discernment}, i.e., the set of mutually exclusive outcomes of the system.
    A \emph{mass function} $m$ assigns masses to all the subsets of $\calY$, $m: 2^\calY \rightarrow [0,1]$, such that $m(\emptyset)=0$ and $\sum_{S \subseteq \calY} m(S)=1$. 
    The value $m(S)$ represents the mass committed \emph{exactly} to the set $S$, without attributing mass to any of its subsets.
    The cardinality of the subset receiving mass reflects the specificity of the available information.
    Mass assigned to singleton sets supports individual outcomes, whereas mass assigned to larger subsets indicates uncertainty regarding which element of the subset is correct.
    Consequently, a shift of mass toward larger subsets corresponds to a less specific representation of the available information.
    In this sense, mass assigned to larger subsets is often associated with epistemic uncertainty arising from a lack of information, while uncertainty distributed across several singleton sets is typically interpreted as reflecting aleatoric uncertainty.
    In particular, assigning mass to the full set $\calY$ represents complete ignorance.

    From the mass function, \emph{belief} and \emph{plausibility} functions are derived as~\cite{shafer1990perspectives}
    \begin{align}
        b(S) &= \sum_{C \subseteq S} m(C), \\
        p(S) &= \sum_{C \cap S \neq \emptyset} m(C), \qquad \forall S \subseteq \calY.
    \end{align}
    
    These functions satisfy $b(S) \leq p(S)$.
    A central operation in Dempster-Shafer theory is \emph{mass fusion}, which provides a mechanism for combining multiple independent sources of evidence.
    Such situations arise naturally when information about the same input is available from multiple sources, through classifiers operating on complementary features or measurements collected by different sensors.
    For example, in autonomous driving one may want to combine information obtained from different camera and radar data.
    In these settings, each source is associated with a mass function, and Dempster-Shafer theory provides dedicated rules for aggregating them.

    Two common fusion rules are the \emph{conjunctive} and \emph{disjunctive} combination rules, which correspond to different assumptions about the reliability and compatibility of the evidence sources~\cite{denoeux2008conjuctive}.
    Given $P$ independent mass functions $\{m_p\}_{p=1}^P$ defined on the same frame of discernment $\calY$, the conjunctive rule is defined as
    \begin{align}
    \label{eq:conjuctiverule}
    m_{\cap}(S)
    &=
    \sum_{\substack{C_1 \cap \dots \cap C_P = S}}
    \prod_{p=1}^P m_p(C_p)
    \qquad \forall\, S \subseteq \calY,
    \end{align}

    where the summation runs over subsets $C_1,\ldots,C_P \subseteq \calY$.
    Here, incompatibilities between evidence sources manifest as mass assigned to the empty set, $m_{\cap}(\emptyset)$, which quantifies conflict between different sources.
    
    The disjunctive rule represents a more cautious fusion strategy, appropriate when at least one evidence source is assumed to be reliable but their mutual consistency is uncertain.
    It is defined as
    \begin{align}
    \label{eq:disjuctiverule}
    m_{\cup}(S)
    &=
    \sum_{\substack{C_1 \cup \dots \cup C_P = S}}
    \prod_{p=1}^P m_p(C_p)
    \qquad \forall\, S \subseteq \calY.
    \end{align}
    
    By construction, this rule avoids explicit conflict and shifts mass toward larger subsets.

\subsection{Dempster-Shafer classifier}\label{s:DSTinML}
    Dempster-Shafer theory has recently been applied in classical machine learning through multi-class classification models that output Dempster-Shafer mass functions over the power set of labels~\cite{kempkes2024reliable}.
    Unlike conventional probabilistic classifiers, these models can assign support not only to individual labels but also to sets of labels.
    In Ref.~\cite{kempkes2024reliable}, this is achieved by learning a full Dempster-Shafer mass function over $2^{\calY}$.
    
    While this formulation enables a rich representation of uncertainty, it incurs exponential scaling in the number of classes.
    To address this challenge, an efficient alternative classifier is proposed which avoids the explicit representation of all $2^n$ masses.
    Instead, it operates on lower-dimensional quantities derived from the mass function.
    Specifically, the model outputs the singleton plausibilities, whose number scales linearly with the number of classes.
    Although the full mass function is not explicitly represented, this approach still supports classification and uncertainty quantification through suitable post-processing procedures.
    
    Let $\hat{p}, p \in [0,1]^n$ denote the vectors of predicted and target singleton plausibilities derived from the ground-truth mass function, with components
    \begin{align} \hat{p}_k \coloneqq \hat p(\{k\}), \qquad p_k \coloneqq p(\{k\}), \qquad k \in [n].
    \end{align}
    The \emph{ground-truth mass function} $m$ encodes the information provided by the class labels.
    For a training example with true label $y \in \calY$, it assigns unit mass to the corresponding singleton and zero mass to all other subsets:
    \begin{align}
        m(\{y\}) = 1, \qquad m(S) = 0,
    \label{eq:true_mass}
    \end{align}
    for all $S \subseteq \calY,\; S \neq \{y\}$.
    The associated target singleton plausibilities therefore satisfy $p_k = \delta_{k,y}$.

    The loss function below encourages the predicted singleton plausibilities to approximate this target while still permitting uncertainty over multiple labels
    \begin{align}
        \mathcal{L}(\hat p, p) = \lambda \|\hat p-p\|_2^2- (1-\lambda)\, p \cdot \log(\hat p),
    \label{eq:pl_sglloss}
    \end{align}
    where $\log(\hat p)$ is applied component-wise.

    The squared-error term encourages the predicted singleton plausibilities to match the target plausibilities and, in particular, penalizes plausibility assigned to incorrect classes.
    In contrast, the cross-entropy term depends only on the plausibility assigned to the correct class.
    Since singleton plausibilities are not required to sum to one, the cross-entropy term alone does not penalize assigning high plausibility to multiple labels simultaneously.
    Consequently, the trade-off parameter $\lambda$ controls the extent to which the model is encouraged to produce specific predictions rather than uncertainty distributed across multiple labels.
    The effect of $\lambda$ is analyzed in more detail in Section~\ref{s:numerics}, with an illustrative example shown in Fig.~\ref{fig:lambda_influence}.

\section{Results}
    
\subsection{Dempster-Shafer classifier with parameterized quantum circuits}\label{s:trainingalgo}
    Consider a multi-class classification task with label set $\calY = \{1, \dots, n\}$.
    Given a training set $ \calD=\{(x_i,y_i)\}_{i=1}^{N}$ consisting of inputs $x_i\in\bbR^d$ and outputs $y_i\in\calY$ drawn i.i.d. from some fixed but unknown distribution, the goal is to learn a hypothesis which maps inputs to outputs.

    To realize such a hypothesis, we employ an $n$-qubit parameterized quantum circuit (PQC), where each class is associated with one qubit.
    For input $x$ and trainable parameters $\vartheta\in\bbR^p$, the PQC prepares the quantum state
    \begin{align}
        \rho(x,\vartheta) = U(x,\vartheta)\rho_0U^\dagger(x,\vartheta),
    \label{eq:rho_pqc}
    \end{align}
    where $\rho_0$ denotes the initial state and $U(x,\vartheta)$ the unitary implemented by the circuit.
    In the following, we drop the dependence of $\rho$ on $x$ and $\vartheta$ for simplicity.
    For each input $x$ and parameters $\vartheta$, the basis state probabilities of $\rho$ define a Dempster-Shafer \emph{mass function} $m_\rho:2^{\calY} \to [0,1]$.
    Specifically, let $i=(i_1,\ldots,i_n)\in\{0,1\}^n$ denote the binary representation of basis state $\ket{i}$.
    We associate to $\ket{i}$ the subset $S_i = \{k\in\calY \mid i_k = 1\}$.    
    The mass assigned to $S_i$ is then given by
    \begin{align}
        m_\rho(S_i) = \Tr\!\left\{\rho(x,\vartheta)\ket{i}\bra{i}\right\}.
    \label{eq:mass_from_rho}
    \end{align}

    \begin{remark}
        The encoding in Eq.~\eqref{eq:mass_from_rho} does not enforce the constraint $m_\rho(\emptyset)=0$.
        Empirically, however, we observe that the learned empty-set mass is small and decreases with the number of qubits (Appendix~\ref{A:expdetails}, Fig.~\ref{fig:emptysetmass}).
    \end{remark}

    We identify subsets $S\subseteq\calY$ with their corresponding bit strings $i\in\{0,1\}^n$ whenever convenient.
    Further, for notational simplicity, we write $m$ instead of $m_\rho$ whenever the underlying state is clear from context.

    In the following proposition, we describe how two other relevant quantities from DST, namely singleton \emph{beliefs} and \emph{plausibilities}, are defined in this setting.

    \begin{proposition}[Singleton belief and plausibility]\label{prop:plbelsgl}
        Let $\rho$ be an $n$-qubit quantum state as in Eq.~\eqref{eq:rho_pqc} and let $m_\rho$ be the mass function induced by Eq.~\eqref{eq:mass_from_rho}.
        For each $k \in \calY$, define the singleton belief $b_k \coloneqq b(\{k\})$ and singleton plausibility $p_k \coloneqq p(\{k\})$.
        Then, these quantities are given by
        \begin{align}
            \label{eq:sglbel}
            b_k &= \Tr\!\left\{\rho \ket{k}\bra{k}\right\}, \\
            \label{eq:sglpl}
            p_k &= \Tr\!\left\{\rho_k (\ket{1}\bra{1})_k\right\},
        \end{align}
        where $\rho_k$ denotes the reduced density matrix of qubit $k$, $(\ket{1}\bra{1})_k$ is the projector onto the state $\ket{1}$ acting on qubit $k$, and $\ket{k}$ denotes the computational basis state with Hamming weight one corresponding to class $k$.
    \end{proposition}

    The proof of Proposition~\ref{prop:plbelsgl} can be found in Appendix~\ref{A:proofs}.
    We define the $k$-th output of the PQC-based hypothesis function as $[f_\vartheta(x)]_k = \Tr\!\left\{\rho_k (\ket{1}\bra{1})_k\right\}$, where $k \in [n]$.
    The predicted class is then given by $\arg\max_{k\in[n]} [f_\vartheta(x)]_k$.   

    Given the loss function in Eq.~\eqref{eq:pl_sglloss}, our goal in multi-class classification is to minimize the expected loss over the problem distribution.
    In practice, the problem distribution is unknown and we explicitly optimize the empirical risk, which is defined as the average loss over the set $\calD$.

    By Proposition~\ref{prop:plbelsgl}, the quantities in the loss function can be expressed as expectation values of single-qubit projectors.
    Therefore, these quantities can be estimated directly from repeated local measurements on the quantum state prepared by the parameterized circuit. 
    Standard concentration bounds, such as Hoeffding’s inequality, imply that all singleton plausibilities can be efficiently estimated on a quantum circuit.
    In particular, achieving accuracy $\epsilon$ and confidence $1-\delta$ requires $\mathcal{O}\!\left(\frac{1}{\epsilon^2}\log\frac{2n}{\delta}\right)$ samples.

    \begin{remark}
        For $\lambda = 1$, the loss in Eq.~\eqref{eq:pl_sglloss} reduces to a mean-squared error on these scores, making this readout a natural choice for multi-class training with parameterized quantum circuits.
    \end{remark}
    
    After training, one may want to approximate the full mass vector.
    In general, explicitly sampling the complete $2^n$-dimensional mass vector is computationally inefficient when the mass function is dense.
    We therefore analyze \emph{when the mass functions exhibit sparsity}, as this would allow for efficient approximation from measurement data.
    
    In particular, we consider an idealized setting in which the singleton plausibilities are binary.
    Then, the corresponding mass vector is uniquely determined, as stated in the following proposition.

    \begin{proposition}[Unique mass for binary singleton plausibility]\label{prop:uniquemass}
        Consider the $n$-dimensional vector of singleton plausibilities $(p_k)_{k=1}^n$. 
        If $p_k\in\{0,1\}$ for all $k$, then the corresponding mass function $m$ over $2^{\calY}$ is unique and given by
        \begin{align}
        m(S) =
        \begin{cases}
        1, & \text{if } S = S^\ast, \\
        0, & \text{otherwise},
        \end{cases}
        \end{align}
        where $S^\ast = \{\, k \in \calY \mid p_k = 1 \,\}$.
    \end{proposition}
    
    The proof is deferred to Appendix~\ref{A:proofs}.
    Proposition~\ref{prop:uniquemass} formalizes the observation that binary singleton plausibilities uniquely determine a $1$-sparse mass function concentrated on a single set $S^\ast$. 
    Consequently, in the idealized limit of zero loss of the singleton plausibility objective (Eq.~\eqref{eq:pl_sglloss}) with binary targets, the learned mass functions are expected to approach this sparse regime.

    For non-binary singleton plausibilities, however, the corresponding mass function is generally not unique.
    In this regime, the learned mass allocation is shaped jointly by the data through the singleton plausibility objective and by the inductive bias of the chosen PQC ansatz and optimization procedure. 
    Proposition~\ref{prop:uniquemass} indicates that this ambiguity progressively diminishes as the predicted plausibilities approach their binary targets.
    
    In practical machine learning scenarios, however, the loss will not be exactly zero and hence predicted singleton plausibilities are generally not binary.
    We therefore investigate empirically how the sparsity of the learned mass functions depends on the quality of the learned plausibility estimates in Section~\ref{s:numerics}, as illustrated in Fig.~\ref{fig:sparsity}.

\subsection{Post-processing of model outputs}\label{s:postprocessing}

    Once the model has been trained, the learned mass function can be used in different post-processing procedures.
    Dempster-Shafer theory provides several ways to quantify uncertainty, which require access to different quantities derived from the mass function in Eq.~\eqref{eq:mass_from_rho} and therefore incur different computational costs.
    In this section, we introduce several post-processing procedures that can be implemented efficiently within the proposed framework.
    All quantities considered in the following can be estimated from measurement outcomes of the PQC.
    Using standard sampling, an additive precision $\epsilon$ can be achieved with sample complexity $\mathcal{O}(\epsilon^{-2})$.
    Alternatively, quantum amplitude estimation yields a quadratic improvement to $\mathcal{O}(\epsilon^{-1})$ at the expense of increased circuit depth.\\
    
    \paragraph{Summed singleton plausibilities.}
    We introduce a simple uncertainty measure obtained by summing the singleton plausibilities associated with the $n$ output classes,
    \begin{align}
        \sigma_p = \sum_{k=1}^n p_k.
    \end{align}
    The intuition is that larger values of $\sigma_p$ indicate that substantial plausibility is assigned to multiple labels simultaneously, reflecting 
    higher overall uncertainty.
    Since $\sigma_p$ depends only on the singleton plausibilities, it can be estimated efficiently from the quantities in Proposition~\ref{prop:plbelsgl} without reconstructing the full mass function.\\

    \paragraph{Volume of singleton credal set.}
    Another measure of uncertainty is the volume of the so-called singleton credal set denoted $\mathcal{C}(m)$.
    This set consists of all probability distributions $q \in \Delta^{n-1}$ compatible with the mass function $m$, where $\Delta^{n-1}$ denotes the $(n-1)$-dimensional probability simplex.
    Formally,
    \begin{align}
        \mathcal{C}(m) = \left\{\, q \in \Delta^{n-1} \;\middle|\; 
        b_k \le q_k \le p_k,\ \forall k \in \calY \,\right\},
    \end{align}
    where $b_k = b(\{k\})$ and $p_k = p(\{k\})$ denote the singleton belief and plausibility values, respectively, and $\calY = \{1,\dots,n\}$. Note that the singleton credal set is a superset of the full credal set obtained by considering belief and plausibility constraints for all subsets.    
    In our setting, singleton beliefs and plausibilities can be estimated efficiently from measurement data, thereby fully specifying the singleton credal set.\\

    \paragraph{Averaged class-wise $(p-b)$-gap.}
    As an alternative uncertainty measure, we introduce the averaged class-wise gap between singleton plausibility and belief, as
    \begin{align}
        g_{p-b} = \frac{1}{n}\sum_{k=1}^n \bigl(p_k - b_k\bigr). 
    \end{align}
    For each class, this quantity measures the width of the probability interval induced by the mass function and averages it across all classes.
    Larger values therefore indicate greater uncertainty.
    Since $g_{p-b}$ depends only on singleton beliefs and plausibilities, it can be estimated efficiently from quantum measurements.\\ 
        
    \begin{figure*}[t]
        \includegraphics[width=\textwidth]{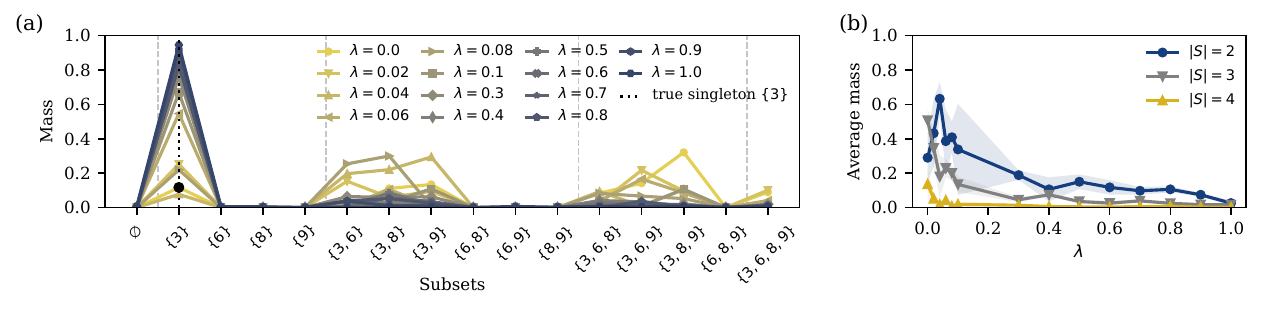}
        \caption{Illustration of the influence of the loss function parameter $\lambda$. (a) Exemplary mass vector for a representative MNIST image of the digit $3$ for different values of $\lambda$. Smaller values of $\lambda$ increase the mass assigned to larger subsets, resulting in more imprecise hypotheses. (b) Average mass assigned to subsets of different cardinalities $|S|$ as a function of $\lambda$, quantitatively confirming the trend observed in panel (a).}
        \label{fig:lambda_influence}
    \end{figure*}
  
    \paragraph{Mass vector fusion.}
    Dempster-Shafer theory provides a principled framework for combining evidence from multiple sources.
    In a machine learning setting, such sources may correspond to independently trained classifiers, models with different architectures, or models trained on different datasets.
    We therefore consider $P \in \mathbb{N}$ independently trained parameterized quantum circuits, each producing a mass function $m_{\rho_p}$ for $p \in [P]$ for a given input.
    The resulting mass functions can then be combined using the fusion rules introduced in Section~\ref{s:DST}.

    Ref.~\cite{luo2026attributefusion} shows that DST fusion operations can be performed directly on quantum-encoded mass functions using polynomial quantum resources.
    This contrasts with naive fusion procedures based on explicitly represented mass vectors, whose dimension scales as $2^n$ with the number of classes.
    Consequently, the quantum fusion procedure may offer benefits in settings where such explicit representations are manipulated.\footnote{This does not establish a provable quantum advantage, as no corresponding classical hardness result is known and more efficient classical algorithms may exist.}

    A key assumption of Ref.~\cite{luo2026attributefusion} is the availability of quantum states encoding the mass functions to be fused.
    Preparing such states is not efficient in general and may itself become a computational bottleneck.
    In our setting, this assumption is naturally satisfied: the mass functions arise directly as measurement statistics of trained PQCs.
    Moreover, quantities of interest after fusion, such as singleton beliefs and plausibilities, can be estimated directly from the fused quantum state without reconstructing the fused, potentially dense, mass vector.
    Our framework therefore provides a practical setting in which quantum DST fusion procedures of Ref.~\cite{luo2026attributefusion} can be applied to learned mass functions.
   
\subsection{Numerical results}\label{s:numerics}
    In this section, we present numerical experiments that illustrate and complement the theoretical results of the previous section.
    We first investigate how the loss function parameter $\lambda$ controls the uncertainty behavior of the proposed Dempster-Shafer quantum classifier on an exemplary classification problem.
    We then empirically examine to what extent the sparsity of the learned mass function, shown in Proposition~\ref{prop:uniquemass} under idealized assumptions, persists in practical settings.
    Finally, we evaluate different uncertainty measures from Section~\ref{s:postprocessing} on a realistic image-classification task and demonstrate how Dempster-Shafer mass fusion can be used to aggregate evidence from multiple classifiers to improve predictive performance. Further details on the numerical experiments are provided in Appendix~\ref{A:expdetails}.\\
    
    \paragraph{Influence of the loss function parameter $\lambda$.}
    To illustrate the influence of the loss function parameter $\lambda$, we train a quantum classifier on four MNIST digits for varying values of $\lambda$. Figure~\ref{fig:lambda_influence} (a) shows an exemplary mass function for an image of the digit $3$.
    As $\lambda$ decreases, mass is shifted from the correct singleton $\{3\}$ toward larger subsets containing the true label, resulting in increasingly uncertain predictions.
    At the same time, subsets that do not contain the true label receive negligible mass.
    This indicates that the classifier has correctly identified implausible labels while expressing increasing uncertainty about the true class.
    Figure~\ref{fig:lambda_influence} (b) quantifies this behavior by showing the average mass assigned to subsets of different cardinalities $|S|$ as a function of $\lambda$.
    The mass assigned to larger subsets decreases as $\lambda$ increases, confirming the trend observed in panel (a).
    We therefore interpret $\lambda$ as a hyperparameter that controls the caution of the resulting predictions.
    Smaller values lead to more imprecise but uncertainty-aware predictions, while larger values favor concentrated mass assignments.
    The parameter $\lambda$ can easily be optimized through standard hyperparameter search procedures.

    \begin{figure*}[t]
        \includegraphics[width=\textwidth]{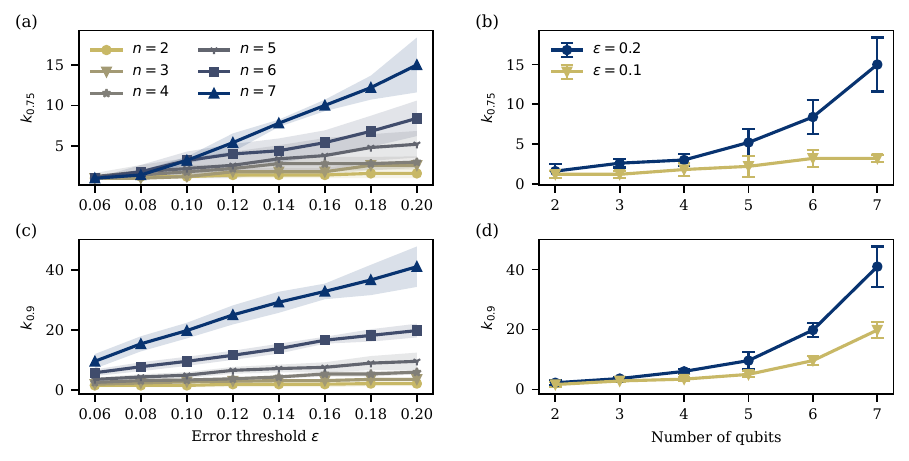}
        \caption{Empirical sparsity of the learned mass vector beyond the ideal setting of Proposition~\ref{prop:uniquemass}. The quantity $k_\tau$ denotes the minimum number of subsets required to account for a fraction $\tau$ of the total mass. Panels (a) and (c) show $k_\tau$ as a function of the error threshold $\varepsilon$ for different numbers of qubits, while panels (b) and (d) show the dependence on the number of qubits for fixed values of $\varepsilon$. As the training error increases, the mass functions become less sparse. Moreover, for fixed error thresholds, $k_\tau$ grows rapidly with the number of qubits.}
        \label{fig:sparsity}
    \end{figure*}

    \paragraph{Empirical sparsity of learned mass functions.}
    Proposition~\ref{prop:uniquemass} establishes that binary singleton plausibilities uniquely determine a sparse mass function.
    Since the loss in Eq.~\eqref{eq:pl_sglloss} encourages singleton plausibilities to approach binary target values, it is natural to ask whether the resulting mass function is also sparse if the loss is not exactly zero.
    We therefore investigate empirically how the sparsity of the learned mass functions depends on the deviation of the singleton plausibilities from their binary targets.
    
    For this purpose, we train PQCs with varying numbers of qubits $n$ on MNIST classification tasks.
    Training is terminated once the mean absolute singleton plausibility error falls below a prescribed threshold $\varepsilon$, defined as $\varepsilon = \frac{1}{Nn}\sum_{i=1}^{N}\sum_{k=1}^{n} \left|
    \hat{p}_k - p_k\right|$, where $N$ is the number of train samples.
    Hence, $\varepsilon$ measures the average deviation from the ideal binary prediction for which Proposition~\ref{prop:uniquemass} guarantees a sparse mass function.

    To quantify sparsity, we consider the quantity $k_\tau$, defined $k_\tau \coloneqq \min\Bigl\{|C| : C \subseteq 2^{\mathcal Y}, \sum_{S \in C} m(S) \ge \tau \Bigr\}$, where $|C|$ denotes the number of subsets contained in the collection $C$.
    Thus, $k_\tau$ is the minimum number of sets whose combined mass is at least $\tau$ and
    smaller values of $k_\tau$ indicate sparser mass functions. 

    Several trends can be observed.
    From Figure~\ref{fig:sparsity} (a) and (c) we see that $k_\tau$ increases with the error threshold $\varepsilon$, indicating that the mass functions become less sparse as the singleton plausibilities deviate further from their binary targets. 
    From panels (b) and (d) one can see that the growth of $k_\tau$ becomes more pronounced as the number of qubits increases.
    For sufficiently small systems, a large fraction of the mass remains concentrated on only a few subsets.
    However, for larger values of $n$, increasingly many subsets are required to capture the same fraction of the total mass.
    This effect is particularly visible in Figures~\ref{fig:sparsity}(b) and (d), where $k_\tau$ grows rapidly with the number of qubits.

    These experiments suggest that effective sparsity can be expected only when the classifier is trained to high precision and when a moderate fraction of the total mass suffices for the intended application.
    Consequently, the exact reconstruction of a complete mass vector may become impractical for larger classification tasks.
    We therefore turn to the analysis of post-processing procedures that can be applied without explicitly reconstructing the full mass vector.

    \paragraph{Out-of-distribution detection with uncertainty metrics.}
    To evaluate the proposed uncertainty measures, we consider an out-of-distribution (o.o.d.) detection task.
    The underlying assumption is that predictions on o.o.d. samples should have higher uncertainty than predictions on in-distribution (i.d.) samples.
    Consequently, a useful uncertainty measure should be able to distinguish between i.d. and o.o.d. data.
    We quantify this ability using the area under the receiver operating characteristic curve (AUROC), where larger values indicate better separation between the two distributions.
    
    In the first experiment, a parameterized quantum circuit with $10$ qubits is trained on MNIST using the loss function in Eq.~\eqref{eq:pl_sglloss}. The trained model is then evaluated on MNIST test samples (i.d.) and FashionMNIST~\cite{fashionMNISt} samples (o.o.d.). The resulting AUROC values are reported in Table~\ref{tab:auroc_ood}. The summed singleton plausibility and the average plausibility-belief gap achieve AUROC values of $0.83$ and $0.82$, respectively, indicating good discrimination between i.d. and o.o.d. samples. In contrast, the singleton credal set volume performs close to random guessing with an AUROC of $0.54$.    
    A second experiment is conducted using CIFAR10 as the in-distribution dataset and FashionMNIST as the out-of-distribution dataset, where similar qualitative behavior is observed.

    \begin{table}[t]
        \centering
        \begin{tabular}{l cc}
            \toprule
            \textbf{Uncertainty metric} & \textbf{MNIST} & \textbf{CIFAR10} \\
            \midrule
            Summed singleton plausibility & $0.83$ & $0.707$ \\
            Average plausibility-belief gap & $0.82$ & $0.684$ \\
            Credal set volume & $0.54$ & $0.625$ \\
            \bottomrule
        \end{tabular}
        \vspace{1\baselineskip}
        \caption{AUROC values for out-of-distribution detection on two datasets using different uncertainty metrics derived from the Dempster-Shafer quantum classifier. All metrics are defined in Section~\ref{s:postprocessing}.}
        \label{tab:auroc_ood}
    \end{table}

\begin{figure}[t]
        \includegraphics[width=1\columnwidth]{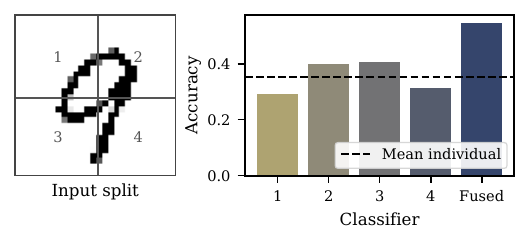}
        \caption{Mass fusion improves classification performance. Four individual classifiers are trained on disjoint patches of MNIST images, obtained by splitting each image into four parts, and evaluated separately. The fused classifier combines learned mass functions using Dempster’s rule as described in Section~\ref{s:postprocessing}. Bars show test accuracy of each individual classifier and of the fused model; the dashed line indicates the mean accuracy of the individual classifiers.}
        \label{fig:fused_classifier}
    \end{figure}

    \paragraph{Mass fusion from multiple classifiers.}
    Beyond assessing the uncertainty of individual predictions, Dempster-Shafer theory also provides an approach for evidence aggregation from multiple models.
    Figure~\ref{fig:fused_classifier} demonstrates how mass fusion can be used to aggregate evidence from multiple quantum classifiers trained on complementary data representations.
    
    In this experiment, each PQC classifier with $10$ qubits is trained on a different patch of the MNIST input images, obtained by partitioning each image into four disjoint quadrants.
    While each individual model achieves only moderate accuracy when operating on partial information, combining their learned mass functions using Dempster’s disjunctive rule (Eq.~\eqref{eq:disjuctiverule}) leads to a clear improvement in overall classification performance.
    
    As shown in Fig.~\ref{fig:fused_classifier}, the accuracy of the fused classifier, computed on the singleton plausibilities, consistently exceeds that of each individual model as well as their mean performance.
    This demonstrates how Dempster-Shafer–based fusion can enable aggregation of complementary evidence sources, resulting in potentially more precise predictions than those obtained from any single classifier alone.

    In the present experiments, the fusion step is performed classically, which is feasible for small problem sizes considered here.
    For larger problems, the exponentially growing number of possible subsets may pose computational challenges for classical fusion.
    Ref.~\cite{luo2026attributefusion} showed that DST fusion can be performed directly on quantum-encoded mass functions using only polynomial quantum resources.
    This approach assumes access to quantum states encoding the mass functions to be fused.
    As discussed in Section~\ref{s:postprocessing}, our framework naturally provides such quantum representations through the learned parameterized quantum circuits.
    Moreover, our framework and the proposed post-processing procedures enable efficient readout for both classification and uncertainty quantification.

\section{Discussion}
    In this work, we address the problem of \emph{uncertainty quantification} in quantum machine learning.
    Quantifying predictive uncertainty is an important step towards practical quantum machine learning, particularly in applications where decisions must be accompanied by confidence assessments.
    To this end, we introduce a classification framework based on Dempster-Shafer theory (DST) for parameterized quantum circuits (PQCs).

    The proposed framework establishes a correspondence between quantities from DST and measurement outcomes of PQCs.
    In particular, so-called \emph{mass functions} are represented through basis state probabilities of the quantum state.
    Building on a training strategy from classical machine learning~\cite{kempkes2024reliable}, we show that the resulting loss function can be evaluated efficiently from measurement data.
    We further show analytically that, in the ideal setting of zero loss, the learned quantum state is associated with a sparse mass function.
    In numerical experiments, we investigate to what extent this sparsity persists under more realistic training conditions.
    In addition, we introduce several post-processing procedures that can be estimated efficiently from measurement data and provide uncertainty information for individual predictions.
    
    We illustrate the versatility of the framework through several numerical experiments, in settings such as out-of-distribution detection and fusion of mass functions from multiple sources.
    In the case of mass fusion, our framework may provide an option for efficient state preparation and readout for the quantum fusion procedures introduced in Ref.~\cite{luo2026attributefusion}.
    The learned PQCs directly generate the required quantum-encoded mass representations, and relevant post-processing quantities can be estimated from the fused quantum state without reconstructing the full mass vector.
    This combination provides a direct interface between uncertainty-aware quantum learning models and quantum fusion procedures while avoiding explicit reconstruction of the full mass vector.
    
    Several limitations remain. The numerical experiments are intended as proof-of-concept demonstrations and are restricted to simulated quantum systems.
    A more extensive empirical evaluation, including comparisons with established uncertainty quantification methods, is left for future work.
    Moreover, it remains open to what extent similar uncertainty representations and associated post-processing procedures can be realized efficiently in classical or quantum-inspired models.
    While Ref.~\cite{kempkes2024reliable} shows that singleton plausibilities admit efficient classical representations, it is currently unclear whether an analogous result holds for singleton beliefs.    
    A further limitation concerns the identifiability of the learned mass functions. The singleton plausibility objective in Eq.~\eqref{eq:pl_sglloss} directly constrains only the singleton plausibilities, and non-binary plausibility vectors generally do not uniquely determine the corresponding mass function. Consequently, the learned distribution of mass across subsets of $\calY$ is influenced by both the training data and the inductive bias of the PQC ansatz and optimization procedure.
    While Proposition~\ref{prop:uniquemass} formalizes the observation that this ambiguity vanishes in the binary plausibility limit, investigating additional regularization principles or refinement procedures that further constrain the mass function constitutes an interesting direction for future work.

    Taken together, our results show that Dempster-Shafer theory provides a natural foundation for uncertainty-aware quantum machine learning and represents a step toward more reliable quantum models, as well as a new perspective for analyzing quantum systems beyond standard performance metrics.

\section{Acknowledgments}
    V.D.\ acknowledges support from the European Union’s Horizon Europe program through the ERC CoG BeMAIQuantum (Grant No. 101124342). V.D. and E.vN acknowledge the Dutch National Growth Fund (NGF) as part of the Quantum Delta NL programme.
    GPT‑5.6 was used to assist with code development, writing and mathematical consistency checks during the preparation of this manuscript.
    All scientific ideas, analyses, and conclusions are those of the authors, and all AI-assisted code and text were carefully reviewed, manually checked, and validated prior to inclusion.

\section{Disclaimer}
    The results, opinions, and conclusions expressed in this publication are not necessarily those of Volkswagen Aktiengesellschaft or those of the European Union or the European Research Council.
    None of the parties involved can be held responsible for them.

\bibliography{references}

@article{biamonte2017quantum,
  title={Quantum Machine Learning},
  author={Biamonte, Jacob and Wittek, Peter and Pancotti, Nicola and Rebentrost, Patrick and Wiebe, Nathan and Lloyd, Seth},
  journal={Nature},
  volume={549},
  number={7671},
  pages={195--202},
  year={2017},
  publisher={Nature Publishing Group UK London},
  url ={https://doi.org/10.1038/nature23474},
  doi ={10.1038/nature23474}
}

@article{dunjko2018machine,
  title={Machine Learning \& Artificial Intelligence in the Quantum Domain: A Review of Recent Progress},
  author={Dunjko, Vedran and Briegel, Hans J},
  journal={Reports on Progress in Physics},
  volume={81},
  number={7},
  pages={074001},
  year={2018},
  publisher={IOP Publishing},
  doi = {10.1088/1361-6633/aab406},
  url = {https://dx.doi.org/10.1088/1361-6633/aab406}
}

@article{carleo2019machine,
  title={Machine Learning and the Physical Sciences},
  author={Carleo, Giuseppe and Cirac, Ignacio and Cranmer, Kyle and Daudet, Laurent and Schuld, Maria and Tishby, Naftali and Vogt-Maranto, Leslie and Zdeborov{\'a}, Lenka},
  journal={Reviews of Modern Physics},
  volume={91},
  number={4},
  pages={045002},
  year={2019},
  publisher={APS},
  doi = {10.1103/RevModPhys.91.045002},
  url = {https://link.aps.org/doi/10.1103/RevModPhys.91.045002}
}

@book{schuld2021machine,
  title={Machine Learning with Quantum Computers},
  author={Schuld, Maria and Petruccione, Francesco},
  volume={676},
  year={2021},
  publisher={Springer},
  url ={https://doi.org/10.1007/978-3-030-83098-4}
}

@article{cerezo2021variational,
  title={Variational Quantum Algorithms},
  author={Cerezo, Marco and Arrasmith, Andrew and Babbush, Ryan and Benjamin, Simon C and Endo, Suguru and Fujii, Keisuke and McClean, Jarrod R and Mitarai, Kosuke and Yuan, Xiao and Cincio, Lukasz and others},
  journal={Nature Reviews Physics},
  volume={3},
  number={9},
  pages={625--644},
  year={2021},
  publisher={Nature Publishing Group UK London},
  url ={https://doi.org/10.1038/s42254-021-00348-9},
  doi = {10.1038/s42254-021-00348-9}
}

@article{bharti2022noisy,
  title={Noisy Intermediate-Scale Quantum Algorithms},
  author={Bharti, Kishor and Cervera-Lierta, Alba and Kyaw, Thi Ha and Haug, Tobias and Alperin-Lea, Sumner and Anand, Abhinav and Degroote, Matthias and Heimonen, Hermanni and Kottmann, Jakob S and Menke, Tim and others},
  journal={Reviews of Modern Physics},
  volume={94},
  number={1},
  pages={015004},
  year={2022},
  publisher={APS},
  doi = {10.1103/RevModPhys.94.015004},
  url = {https://link.aps.org/doi/10.1103/RevModPhys.94.015004}
}

@article{liu2021rigorous,
  title={A Rigorous and Robust Quantum Speed-Up in Supervised Machine Learning},
  author={Liu, Yunchao and Arunachalam, Srinivasan and Temme, Kristan},
  journal={Nature Physics},
  volume={17},
  number={9},
  pages={1013--1017},
  year={2021},
  publisher={Nature Publishing Group UK London},
  url = {https://doi.org/10.1038/s41567-021-01287-z},
  doi = {10.1038/s41567-021-01287-z}
}

@article{gyurik2023exponential,
  title={Exponential Separations between Classical and Quantum Learners},
  author={Gyurik, Casper and Dunjko, Vedran},
  journal={arXiv:2306.16028},
  year={2023},
  url={https://arxiv.org/abs/2306.16028}
}

@article{sweke2021quantum,
  title={On the quantum versus classical learnability of discrete distributions},
  author={Sweke, Ryan and Seifert, Jean-Pierre and Hangleiter, Dominik and Eisert, Jens},
  journal={Quantum},
  volume={5},
  pages={417},
  year={2021},
doi={10.22331/q-2021-03-23-417},
  publisher={Verein zur F{\"o}rderung des Open Access Publizierens in den Quantenwissenschaften}
}

@article{pirnay2023superpolynomial,
  title={Superpolynomial quantum-classical separation for density modeling},
  author={Pirnay, Niklas and Sweke, Ryan and Eisert, Jens and Seifert, Jean-Pierre},
  journal={Physical Review A},
  volume={107},
  number={4},
  pages={042416},
  year={2023},
doi={10.1103/PhysRevA.107.042416},
  publisher={APS}
}

@article{huang2021information,
  title={Information-theoretic bounds on quantum advantage in machine learning},
  author={Huang, Hsin-Yuan and Kueng, Richard and Preskill, John},
  journal={Physical Review Letters},
  volume={126},
  number={19},
  pages={190505},
  year={2021},
  publisher={APS},
  doi = {10.1103/PhysRevLett.126.190505},
  url = {https://link.aps.org/doi/10.1103/PhysRevLett.126.190505}
}

@article{molteni2024exponential,
  title={Exponential quantum advantages in learning quantum observables from classical data},
  author={Molteni, Riccardo and Gyurik, Casper and Dunjko, Vedran},
  journal={npj Quantum Information},
  year={2026},
  url={https://doi.org/10.1038/s41534-025-01162-2},
  publisher={Nature Publishing Group UK London}
}

@article{zhou2023bf,
  title={BF-QC: belief functions on quantum circuits},
  author={Zhou, Qianli and Tian, Guojing and Deng, Yong},
  journal={Expert Systems with Applications},
  volume={223},
  pages={119885},
  year={2023},
  publisher={Elsevier},
  url={https://doi.org/10.1016/j.eswa.2023.119885}
}

@article{luo2026attributefusion,
  title={Attribute fusion-based evidential classifier on quantum circuits},
  author={Luo, Hao and Zhou, Qianli and Pan, Lipeng and Li, Zhen and Deng, Yong},
  journal={Quantum Machine Intelligence},
  volume={8},
  number={1},
  pages={51},
  year={2026},
  publisher={Springer},
  url={https://doi.org/10.1007/s42484-026-00395-4}
}

@article{vourdas2014quantum,
  title={Quantum probabilities as Dempster-Shafer probabilities in the lattice of subspaces},
  author={Vourdas, Apostolos},
  journal={Journal of Mathematical Physics},
  volume={55},
  number={8},
  year={2014},
  publisher={AIP Publishing},
  url={https://doi.org/10.1063/1.4891972}
}

@inproceedings{kempkes2024reliable,
  title={Reliable Classifications with Guaranteed Confidence Using the Dempster-Shafer Theory of Evidence},
  author={Kempkes, Marie C and Dunjko, Vedran and van Nieuwenburg, Evert and Spiegelberg, Jakob},
  booktitle={Joint European Conference on Machine Learning and Knowledge Discovery in Databases},
  pages={89--105},
  year={2024},
  organization={Springer},
  url={https://doi.org/10.1007/978-3-031-70344-7_6}
}

@InProceedings{sale2023volumeofcredalset,
  title = 	 {Is the volume of a credal set a good measure for epistemic uncertainty?},
  author =       {Sale, Yusuf and Caprio, Michele and H\"{u}llermeier, Eyke},
  booktitle = 	 {Proceedings of the Thirty-Ninth Conference on Uncertainty in Artificial Intelligence},
  pages = 	 {1795--1804},
  year = 	 {2023},
  editor = 	 {Evans, Robin J. and Shpitser, Ilya},
  volume = 	 {216},
  series = 	 {Proceedings of Machine Learning Research},
  month = 	 {31 Jul--04 Aug},
  publisher =    {PMLR},
  url = 	 {https://proceedings.mlr.press/v216/sale23a.html}
}

@article{mckay1992apractical,
    author = {MacKay, David J. C.},
    title = {Practical Bayesian Framework for Backpropagation Networks},
    journal = {Neural Computation},
    volume = {4},
    number = {3},
    pages = {448-472},
    year = {1992},
    month = {05},
    doi = {10.1162/neco.1992.4.3.448}
}

@inproceedings{lakshminarayanan2017simpleandscalable,
author = {Lakshminarayanan, Balaji and Pritzel, Alexander and Blundell, Charles},
title = {Simple and scalable predictive uncertainty estimation using deep ensembles},
year = {2017},
isbn = {9781510860964},
publisher = {Curran Associates Inc.},
address = {Red Hook, NY, USA},
booktitle = {Proceedings of the 31st International Conference on Neural Information Processing Systems},
pages = {6405–6416},
numpages = {12},
location = {Long Beach, California, USA},
series = {NIPS'17},
url={https://dl.acm.org/doi/10.5555/3295222.3295387}
}

@inproceedings{gal2016dropout,
    author = {Gal, Yarin and Ghahramani, Zoubin},
    title = {Dropout as a Bayesian Approximation: Representing Model Uncertainty in Deep Learning},
    year = {2016},
    publisher = {JMLR.org},
    booktitle = {Proceedings of the 33rd International Conference on International Conference on Machine Learning - Volume 48},
    pages = {1050–1059},
    numpages = {10},
    url={https://dl.acm.org/doi/10.5555/3045390.3045502},
    location = {New York, NY, USA}
}

@article{manchingal2022epistemic,
  title={Epistemic Deep Learning},
  author={Manchingal, Shireen Kudukkil and Cuzzolin, Fabio},
  journal={arXiv preprint arXiv:2206.07609},
  year={2022},
  url={https://arxiv.org/abs/2206.07609} 
}

@article{manchingal2023random,
  title={Random-Set Convolutional Neural Network (RS-CNN) for Epistemic Deep Learning},
  author={Manchingal, Shireen Kudukkil and Mubashar, Muhammad and Wang, Kaizheng and Shariatmadar, Keivan and Cuzzolin, Fabio},
  journal={arXiv preprint arXiv:2307.05772},
  year={2023},
  url={https://arxiv.org/abs/2307.05772} 
}

@article{dempster1967amultivalued,
     ISSN = {00034851},
     author = {A. P. Dempster},
     journal = {The Annals of Mathematical Statistics},
     number = {2},
     pages = {325--339},
     publisher = {Institute of Mathematical Statistics},
     title = {Upper and Lower Probabilities Induced by a Multivalued Mapping},
     urldate = {2023-09-02},
     volume = {38},
     year = {1967},
     url={https://doi.org/10.1007/978-3-540-44792-4_3}
}

@article{shafer1990perspectives,
    title = {Perspectives on the Theory and Practice of Belief Functions},
    journal = {International Journal of Approximate Reasoning},
    volume = {4},
    number = {5},
    pages = {323-362},
    year = {1990},
    issn = {0888-613X},
    doi = {https://doi.org/10.1016/0888-613X(90)90012-Q},
    author = {Glenn Shafer}
}

@article{MNIST,
  title={Gradient-based learning applied to document recognition},
  author={LeCun, Yann and Bottou, Léon and Bengio, Yoshua and Haffner, Patrick},
  journal={Proceedings of the IEEE},
  volume={86},
  number={11},
  pages={2278--2324},
  year={1998},
  publisher={IEEE},
  url={https://ieeexplore.ieee.org/document/726791}
}

@article{fashionMNISt,
  author    = {Han Xiao and
               Kashif Rasul and
               Roland Vollgraf},
  title     = {Fashion-MNIST: a Novel Image Dataset for Benchmarking Machine Learning
               Algorithms},
  journal   = {CoRR},
  volume    = {abs/1708.07747},
  year      = {2017},
  url       = {http://arxiv.org/abs/1708.07747},
  archivePrefix = {arXiv},
  eprint    = {1708.07747},
  bibsource = {dblp computer science bibliography, https://dblp.org}
}

@TECHREPORT{CIFAR10,
  title={Learning Multiple Layers of Features from Tiny Images},
  author={Alex Krizhevsky},
  year={2009},
  url={https://api.semanticscholar.org/CorpusID:18268744}
}

@misc{pennylane,
      title={PennyLane: Automatic differentiation of hybrid quantum-classical computations}, 
      author={Ville Bergholm and Josh Izaac and Maria Schuld and Christian Gogolin and Shahnawaz Ahmed and Vishnu Ajith and M. Sohaib Alam and Guillermo Alonso-Linaje and B. AkashNarayanan and Ali Asadi and Juan Miguel Arrazola and Utkarsh Azad and Sam Banning and Carsten Blank and Thomas R Bromley and Benjamin A. Cordier and Jack Ceroni and Alain Delgado and Olivia Di Matteo and Amintor Dusko and Tanya Garg and Diego Guala and Anthony Hayes and Ryan Hill and Aroosa Ijaz and Theodor Isacsson and David Ittah and Soran Jahangiri and Prateek Jain and Edward Jiang and Ankit Khandelwal and Korbinian Kottmann and Robert A. Lang and Christina Lee and Thomas Loke and Angus Lowe and Keri McKiernan and Johannes Jakob Meyer and J. A. Montañez-Barrera and Romain Moyard and Zeyue Niu and Lee James O'Riordan and Steven Oud and Ashish Panigrahi and Chae-Yeun Park and Daniel Polatajko and Nicolás Quesada and Chase Roberts and Nahum Sá and Isidor Schoch and Borun Shi and Shuli Shu and Sukin Sim and Arshpreet Singh and Ingrid Strandberg and Jay Soni and Antal Száva and Slimane Thabet and Rodrigo A. Vargas-Hernández and Trevor Vincent and Nicola Vitucci and Maurice Weber and David Wierichs and Roeland Wiersema and Moritz Willmann and Vincent Wong and Shaoming Zhang and Nathan Killoran},
      year={2022},
      eprint={1811.04968},
      archivePrefix={arXiv},
      primaryClass={quant-ph},
      url={https://arxiv.org/abs/1811.04968}, 
}

@misc{hewage2026quantummeasurementstatisticsbayesian,
      title={Quantum Measurement Statistics as Bayesian Uncertainty Estimators for Physics-Constrained Learning}, 
      author={Prasad Nimantha Madusanka Ukwatta Hewage and Midhun Chakkravarthy and Ruvan Kumara Abeysekara},
      year={2026},
      eprint={2604.10896},
      archivePrefix={arXiv},
      primaryClass={quant-ph},
      url={https://arxiv.org/abs/2604.10896}, 
}

@article{kempkes2026detecting,
    doi = {10.1088/2632-2153/ae304e},
    url = {https://doi.org/10.1088/2632-2153/ae304e},
    year = {2026},
    month = {jan},
    publisher = {IOP Publishing},
    volume = {7},
    number = {1},
    pages = {015010},
    author = {Kempkes, Marie and Spiegelberg, Jakob and van Nieuwenburg, Evert and Dunjko, Vedran},
    title = {Detecting underdetermination in parameterized quantum circuits},
    journal = {Machine Learning: Science and Technology}
}

@ARTICLE{park2024quantumCP,
  author={Park, Sangwoo and Simeone, Osvaldo},
  journal={IEEE Transactions on Quantum Engineering}, 
  title={Quantum Conformal Prediction for Reliable Uncertainty Quantification in Quantum Machine Learning}, 
  year={2024},
  volume={5},
  number={},
  pages={1-24},
  doi={10.1109/TQE.2023.3333224}
}

@INPROCEEDINGS{wendlinger2025oldrules,
  author={Wendlinger, Maximilian and Tscharke, Kilian and Debus, Pascal},
  booktitle={2025 IEEE International Conference on Quantum Computing and Engineering (QCE)}, 
  title={Old Rules in a New Game: Mapping Uncertainty Quantification to Quantum Machine Learning}, 
  year={2025},
  volume={01},
  number={},
  pages={1803-1814},
  doi={10.1109/QCE65121.2025.00198}
  }

@INPROCEEDINGS{bao2021evidential,
  author={Bao, Wentao and Yu, Qi and Kong, Yu},
  booktitle={2021 IEEE/CVF International Conference on Computer Vision (ICCV)}, 
  title={Evidential Deep Learning for Open Set Action Recognition}, 
  year={2021},
  volume={},
  number={},
  pages={13329-13338},
  doi={10.1109/ICCV48922.2021.01310}}

@article{charpentier2020posterior,
 author = {Charpentier, Bertrand and Z\"{u}gner, Daniel and G\"{u}nnemann, Stephan},
 journal = {Advances in Neural Information Processing Systems},
 editor = {H. Larochelle and M. Ranzato and R. Hadsell and M.F. Balcan and H. Lin},
 pages = {1356--1367},
 publisher = {Curran Associates, Inc.},
 title = {Posterior Network: Uncertainty Estimation without OOD Samples via Density-Based Pseudo-Counts},
 url = {https://proceedings.neurips.cc/paper_files/paper/2020/file/0eac690d7059a8de4b48e90f14510391-Paper.pdf},
 volume = {33},
 year = {2020}
}

@article{sensoy2018evidential,
     author = {Sensoy, Murat and Kaplan, Lance and Kandemir, Melih},
     journal = {Advances in Neural Information Processing Systems},
     editor = {S. Bengio and H. Wallach and H. Larochelle and K. Grauman and N. Cesa-Bianchi and R. Garnett},
     pages = {},
     publisher = {Curran Associates, Inc.},
     title = {Evidential Deep Learning to Quantify Classification Uncertainty},
     url = {https://proceedings.neurips.cc/paper_files/paper/2018/file/a981f2b708044d6fb4a71a1463242520-Paper.pdf},
     volume = {31},
     year = {2018}
}

@article{xiao2026adaptive,
 author = {Xiao, Fuyuan and Zhou, Yu and Pedrycz, Witold},
 journal = {Advances in Neural Information Processing Systems},
 editor = {D. Belgrave and C. Zhang and H. Lin and R. Pascanu and P. Koniusz and M. Ghassemi and N. Chen},
 pages = {17848--17881},
 publisher = {Curran Associates, Inc.},
 title = {An Adaptive Quantum Circuit of Dempster\textquotesingle s Rule of Combination for Uncertain Pattern Classification},
 url = {https://proceedings.neurips.cc/paper_files/paper/2025/file/1a00115ab8033c89c01f3377bc322d9b-Paper-Conference.pdf},
 volume = {38},
 year = {2025}
}

@article{luo2024variational,
  title={Variational quantum linear solver-based combination rules in Dempster--Shafer theory},
  author={Luo, Hao and Zhou, Qianli and Li, Zhen and Deng, Yong},
  journal={Information Fusion},
  volume={102},
  pages={102070},
  year={2024},
  publisher={Elsevier},
  url={https://doi.org/10.1016/j.inffus.2023.102070}
}

@article{mcclean2018barren,
  title={Barren Plateaus in Quantum Neural Network Training Landscapes},
  author={McClean, Jarrod R and Boixo, Sergio and Smelyanskiy, Vadim N and Babbush, Ryan and Neven, Hartmut},
  journal={Nature communications},
  volume={9},
  number={1},
  pages={4812},
  year={2018},
  publisher={Nature Publishing Group UK London},
  url={https://doi.org/10.1038/s41467-018-07090-4}
}

@article{cerezo2025does,
  title={Does provable absence of barren plateaus imply classical simulability?},
  author={Cerezo, Marco and Larocca, Martin and Garc{\'\i}a-Mart{\'\i}n, Diego and Diaz, Nelson L and Braccia, Paolo and Fontana, Enrico and Rudolph, Manuel S and Bermejo, Pablo and Ijaz, Aroosa and Thanasilp, Supanut and others},
  journal={Nature Communications},
  volume={16},
  number={1},
  pages={7907},
  url={https://doi.org/10.1038/s41467-025-63099-6},
  year={2025},
  publisher={Nature Publishing Group UK London}
}

@inproceedings{anschuetz2025unified,
  title={A unified theory of quantum neural network loss landscapes},
  author={Anschuetz, Eric},
  booktitle={International Conference on Learning Representations},
  volume={2025},
  url = {https://proceedings.iclr.cc/paper_files/paper/2025/file/f3680ca31afcb9076846950dbc72c7e3-Paper-Conference.pdf},
  pages={97859--97918},
  year={2025}
}

@inproceedings{gil2025relation,
  title={On the relation between trainability and dequantization of variational quantum learning models},
  author={Gil-Fuster, Elies and Gyurik, Casper and P{\'e}rez-Salinas, Adri{\'a}n and Dunjko, Vedran},
  booktitle={International Conference on Learning Representations},
  volume={2025},
  url = {https://proceedings.iclr.cc/paper_files/paper/2025/file/3c215225324f9988858602dc92219615-Paper-Conference.pdf},
  pages={24069--24093},
  year={2025}
}

@article{gil2024understanding,
  title={Understanding quantum machine learning also requires rethinking generalization},
  author={Gil-Fuster, Elies and Eisert, Jens and Bravo-Prieto, Carlos},
  journal={Nature Communications},
  volume={15},
  number={1},
  pages={2277},
  url={https://doi.org/10.1038/s41467-024-45882-z},
  year={2024},
  publisher={Nature Publishing Group UK London}
}

@misc{barthe2025quantum,
      title={Quantum Advantage in Learning Quantum Dynamics via Fourier coefficient extraction}, 
      author={Alice Barthe and Mahtab Yaghubi Rad and Michele Grossi and Vedran Dunjko},
      year={2025},
      eprint={2506.17089},
      archivePrefix={arXiv},
      primaryClass={quant-ph}
}

@article{hullermeier2021aleatoric,
  title={Aleatoric and {E}pistemic {U}ncertainty in {M}achine {L}earning: {A}n {I}ntroduction to {C}oncepts and {M}ethods},
  author={H{\"u}llermeier, Eyke and Waegeman, Willem},
  journal={Machine Learning},
  volume={110},
  pages={457--506},
  year={2021},
  publisher={Springer}
}

@article{lasserre2015volume,
    title={Volume of slices and sections of the simplex in closed form},
    author={Lasserre, Jean B},
    journal={Optimization Letters},
    volume={9},
    number={7},
    pages={1263--1269},
    year={2015},
    publisher={Springer},
    url={https://doi.org/10.1007/s11590-015-0898-z}
}

@inproceedings{bengs2022pitfalls,
    author = {Bengs, Viktor and H\"{u}llermeier, Eyke and Waegeman, Willem},
    booktitle = {Advances in Neural Information Processing Systems},
    doi = {10.52202/068431-2118},
    editor = {S. Koyejo and S. Mohamed and A. Agarwal and D. Belgrave and K. Cho and A. Oh},
    pages = {29205--29216},
    publisher = {Curran Associates, Inc.},
    title = {Pitfalls of Epistemic Uncertainty Quantification through Loss Minimisation},
    url = {https://proceedings.neurips.cc/paper_files/paper/2022/file/bc1d640f841f752c689aae20b31198c1-Paper-Conference.pdf},
    volume = {35},
    year = {2022}
}

@article{denoeux2008conjuctive,
    title = {Conjunctive and disjunctive combination of belief functions induced by nondistinct bodies of evidence},
    journal = {Artificial Intelligence},
    volume = {172},
    number = {2},
    pages = {234-264},
    year = {2008},
    issn = {0004-3702},
    doi = {https://doi.org/10.1016/j.artint.2007.05.008},
    url = {https://www.sciencedirect.com/science/article/pii/S0004370207001063},
    author = {Thierry Denœux}
}

\vspace*{0.5cm}

\onecolumngrid
\appendix
\renewcommand{\thefigure}{A\arabic{figure}}
\newpage

\setcounter{figure}{0}
\setcounter{theorem}{0}

\begin{center}
\large{Appendix for ``Uncertainty-aware quantum machine learning based on Dempster-Shafer theory''
}
\end{center}

\section{Related work}\label{s:relatedwork}
    \paragraph{Uncertainty quantification in classical machine learning.}
    
    Uncertainty quantification has been extensively studied in classical machine learning, with a wide range of approaches proposed to estimate uncertainty.
    Bayesian neural networks provide a principled probabilistic framework by placing distributions over model parameters, but typically rely on techniques that are computationally expensive~\cite{mckay1992apractical}.
    Ensemble methods approximate predictive uncertainty by analyzing the outputs of multiple independently trained models.
    While empirically effective, this approach incurs a substantial computational overhead during both training and inference~\cite{lakshminarayanan2017simpleandscalable}.
    Monte Carlo dropout offers a lightweight alternative by interpreting dropout as approximate Bayesian inference~\cite{gal2016dropout}, though the resulting uncertainty estimates can be sensitive to architectural and training choices~\cite{lakshminarayanan2017simpleandscalable}.
    Evidential deep learning approaches have gained increasing attention by modeling uncertainty through higher-order distributions over predictive class probabilities~\cite{sensoy2018evidential, bao2021evidential, charpentier2020posterior}. 
    These methods learn such higher-order representations, for instance in the form of Dirichlet distributions or evidential mass assignments~\cite{manchingal2022epistemic, manchingal2023random, kempkes2024reliable}.
    Recent work has highlighted that uncertainty measures learned through empirical risk minimization in evidential or second-order predictive models should be interpreted with care when relating them to epistemic uncertainty~\cite{bengs2022pitfalls}.
    More recently, Ref.~\cite{sale2023volumeofcredalset} showed that the volume of a credal set is not, in general, a reliable measure of epistemic uncertainty for multiclass problems.
    
    \paragraph{Uncertainty in quantum machine learning.}
    In contrast to the classical setting, uncertainty quantification in quantum machine learning has received comparatively limited attention.    
    Several approaches adapt classical uncertainty quantification techniques to the quantum setting. Ref.~\cite{wendlinger2025oldrules} provides a systematic mapping of a range of classical uncertainty quantification methods to quantum machine learning models and investigates their applicability in practice.
    Dempster-Shafer-theoretic approaches, which form the basis of our work, are not considered in that analysis.
    Ref.~\cite{park2024quantumCP} extends conformal prediction to quantum models by adapting this classical framework to account for the stochasticity of quantum measurements and hardware noise.
    Along similar lines, Ref.~\cite{kempkes2026detecting} studies model reliability by transferring the classical notion of underdetermination to parameterized quantum circuits, allowing for the identification of inputs for which predictions are inherently ambiguous.
    Ref.~\cite{hewage2026quantummeasurementstatisticsbayesian} connects quantum measurement statistics to Bayesian uncertainty, showing that repeated circuit evaluations can be interpreted in terms of calibrated posterior uncertainty without requiring explicit probabilistic models.
    
    Connections between Dempster-Shafer theory and quantum mechanics have recently been explored, highlighting conceptual and mathematical similarities between quantum probabilities and mass functions~\cite{vourdas2014quantum}.
    More recently, quantum-inspired and hybrid approaches have studied how mass function representations can be encoded into quantum states and manipulated using quantum operations~\cite{zhou2023bf, luo2024variational, luo2026attributefusion, xiao2026adaptive}.
    These works enable implementations of Dempster–Shafer operations at the level of quantum states, avoiding the explicit manipulation of exponentially large mass vectors required by standard classical implementations.
    However, they generally assume access to quantum representations of masses and do not address how they can be efficiently constructed.

\section{Proof of Proposition~\ref{prop:plbelsgl} and Proposition~\ref{prop:uniquemass}}\label{A:proofs}

    In this section, we provide the proofs of  Proposition~\ref{prop:plbelsgl} and Proposition~\ref{prop:uniquemass}.
    
    \begin{customproposition}{\ref{prop:plbelsgl}}[Singleton belief and plausibility]
        Let $\rho$ be an $n$-qubit quantum state as in Eq.~\eqref{eq:rho_pqc} and let $m_\rho$ be the mass function induced by Eq.~\eqref{eq:mass_from_rho}.
        For each $k \in \calY$, define the singleton belief $b_k \coloneqq b(\{k\})$ and singleton plausibility $p_k \coloneqq p(\{k\})$.
        Then, these quantities are given by
        \begin{align}
            \label{eq:sglbel}
            b_k &= \Tr\!\left\{\rho \ket{k}\bra{k}\right\}, \\
            \label{eq:sglpl}
            p_k &= \Tr\!\left\{\rho_k (\ket{1}\bra{1})_k\right\},
        \end{align}
        where $\rho_k$ denotes the reduced density matrix of qubit $k$, $(\ket{1}\bra{1})_k$ is the projector onto the state $\ket{1}$ acting on qubit $k$, and $\ket{k}$ denotes the computational basis state with Hamming weight one corresponding to class $k$.
    \end{customproposition}

    \begin{proof}[Proof of Proposition~\ref{prop:plbelsgl}]
        We start from the definition of belief in Dempster-Shafer theory.
        For a singleton set $\{k\}$, the belief reduces to the mass assigned exactly to that singleton, which yields
        \begin{align}
            b_k =\sum_{S \subseteq \{k\}} m(S) = m(\{k\}) = \operatorname{Tr}\!\bigl(\rho \ket{k}\bra{k}\bigr),
        \end{align}
        
        establishing Eq.~\eqref{eq:sglbel}.
        Similarly, by definition, the plausibility of the singleton $\{k\}$ is given by the total mass of all subsets that contain $k$,
        \begin{align}
            p_k = p(\{k\}) &= \sum_{S \cap \{k\} \neq \emptyset} m(S).
        \end{align}
        Using the correspondence between subsets and computational basis states, this sum ranges over all bit strings $i \in \{0,1\}^n$ such that $i_k = 1$, yielding
        \begin{align}
            p_k &= \sum_{\substack{i \in \{0,1\}^n \\ i_k = 1}} \bra{i}\rho\ket{i}
            = \operatorname{Tr}\!\bigl(\rho_k \ket{1}\bra{1}_k\bigr),
        \end{align}
        which proves Eq.~\eqref{eq:sglpl}.
    \end{proof}

    \begin{customproposition}{\ref{prop:uniquemass}}[Unique mass for binary singleton plausibility]
        Consider the $n$-dimensional vector of singleton plausibilities $(p_k)_{k=1}^n$. 
        If $p_k\in\{0,1\}$ for all $k$, then the corresponding mass function $m$ over $2^{\calY}$ is unique and given by
        \begin{align}
        m(S) =
        \begin{cases}
        1, & \text{if } S = S^\ast, \\
        0, & \text{otherwise},
        \end{cases}
        \end{align}
        where
        \begin{align}
        S^\ast = \{\, k \in \calY \mid p_k = 1 \,\}.
        \end{align}
    \end{customproposition}

    \begin{proof} 
        Let $S^\ast=\{k\in\calY \mid p_k=1\}$. Since $p_k\in\{0,1\}$ for all $k$, it follows that $p_k=0$ whenever $k\notin S^\ast$.
        
        Consider any $k\notin S^\ast$. Since $p_k = \sum_{S\subseteq\calY:\,k\in S} m(S) = 0$, non-negativity of the masses implies $m(S)=0$ for every subset $S$ containing $k$.
        Hence, every subset $S$ satisfying $S\cap(\calY\setminus S^\ast)\neq\emptyset$ has zero mass.

        Now consider any $k\in S^\ast$ for which we have $p_k=\sum_{S\subseteq\calY:\,k\in S} m(S)=1$.
        Since the masses satisfy $\sum_{S\subseteq\calY} m(S)=1$ and $\sum_{S\subseteq\calY} m(S)=\sum_{S\subseteq\calY:\,k\in S} m(S)+\sum_{S\subseteq\calY:\,k\notin S} m(S)$, it follows that $\sum_{S\subseteq\calY:\,k\notin S} m(S)=0$.
        By non-negativity of masses, we can conclude that $m(S)=0$ for every subset $S$ not containing $k$.

        Since the above arguments hold for every $k\in\calY$, any subset $S$ with nonzero mass must contain every element of $S^\ast$ and no element of $\calY\setminus S^\ast$. 
        Therefore, the only subset that can have nonzero mass is $S^\ast$.
        By normalization of masses, $m(S^\ast)=1$. Consequently,
        \begin{align}
            m(S)= \begin{cases} 1,& S=S^\ast,\\
            0,& \text{otherwise}, \end{cases}
        \end{align}
        which proves the statement.
    \end{proof}

\section{Experimental details and mass of the empty-set}\label{A:expdetails}

    In this section, we provide information on the details of the numerical experiments from Section~\ref{s:numerics}.
    We further analyze numerically the mass assigned to the empty set and how it scales in the number of qubits.
    
\subsection{Experimental details}

    All experiments are implemented using the PennyLane framework~\cite{pennylane}. 
    In the following, we provide details on the data generation, parameterized quantum circuit (PQC) architecture, and training procedure used in our experiments.\\

    \paragraph{Influence of the loss function parameter $\lambda$.}
    To analyze the impact of the loss parameter $\lambda$, we consider a multi-class classification task on four digits of the MNIST dataset~\cite{MNIST}.
    The input data is first reduced to $4$ dimensions using principal component analysis (PCA), fitted on the training data and then  escaled to the interval $[-\pi,\pi]$.
    The model is implemented as a PQC of $4$ qubits and $10$ layers. 
    Classical input data is encoded into the PQC using the AngleEmbedding routine provided by PennyLane.
    Each layer $k$ of the circuit consists of parameterized single-qubit rotations followed by entangling operations. Concretely, for each qubit $i$ and layer $k$, we apply
    \begin{align}
        \mathrm{RX}(\vartheta_{k i 1})\,\mathrm{RY}(\vartheta_{k i 2})\,\mathrm{RZ}(\vartheta_{k i 3}),
    \end{align}
    followed by controlled-NOT (CNOT) gates between neighboring qubits and subsequent re-uploading using the same encoding routine as before. 
    
    Training is performed using mini-batch stochastic optimization with batch size $1$ over $300$ epochs on 5 training samples. 
    We use the Adam optimizer with learning rate $0.01$ to minimize the loss defined in Eq.~\eqref{eq:pl_sglloss}.\\

    \paragraph{Empirical sparsity of learned mass functions.}    
    To investigate the sparsity properties of the learned mass functions under a more realistic setting, we consider a multi-class classification task based on the MNIST dataset.
    The model operates on $n$ qubits, where $n$ is varied from 2 to 7, with each qubit associated to one class.
    The input data is reduced to $n$ dimensions using principal component analysis (PCA) and subsequently rescaled to the interval $[-\pi, \pi]$ to match the range required for quantum feature encoding.
    
    Classical input data is encoded using the AngleEmbedding routine provided by PennyLane. Each layer $k$ of the circuit consists of parameterized single-qubit rotations followed by entangling operations. Concretely, for each qubit $i$ and layer $k$, we apply
    \begin{align}
    \mathrm{RX}(\vartheta_{k i 1})\,\mathrm{RY}(\vartheta_{k i 2})\,\mathrm{RZ}(\vartheta_{k i 3}),
    \end{align}
    followed by CNOT gates between neighboring qubits and subsequent re-uploading using the same encoding routine as before. The PQC consists of 20 such layers.
    
    Training is performed using mini-batch stochastic optimization with batch size $1$ on $5$ training samples over at most $5000$ epochs or until the required error threshold is reached. We use the Adam optimizer with learning rate $0.01$ and set the loss hyperparameter to $\lambda = 0.9$.\\

    \paragraph{Out-of-distribution detection with uncertainty metrics.}
    We evaluate the proposed uncertainty measures in an out-of-distribution (o.o.d.) detection setting using the MNIST dataset with $n=10$ classes.
    The input data is first reduced to $10$ dimensions using principal component analysis (PCA), normalized to the interval $[-\pi,\pi]$, and subsequently encoded into the quantum circuit.
    The training set consists of $n_{\mathrm{train}} = 300$ samples, while evaluation is performed on $n_{\mathrm{test}} = 300$ in-distribution samples and $n_{\mathrm{ood}} = 300$ out-of-distribution samples.
    
    The model is implemented as a parameterized quantum circuit with $n=10$ qubits and $10$ layers.
    Classical input data is encoded using the AngleEmbedding routine provided by PennyLane. Each layer $k$ of the circuit consists of parameterized single-qubit rotations followed by entangling operations. Concretely, for each qubit $i$ and layer $k$, we apply
    \begin{align}
    \mathrm{RX}(\vartheta_{k i 1})\,\mathrm{RY}(\vartheta_{k i 2})\,\mathrm{RZ}(\vartheta_{k i 3}),
    \end{align}
    followed by CNOT gates between neighboring qubits.
    
    Training is performed using mini-batch stochastic optimization with batch size $16$ over $50$ epochs. We employ the Adam optimizer with learning rate $0.1$ and set the loss parameter to $\lambda = 0.8$.
    
    Experiments on CIFAR data are conducted with $n=5$ qubits and $5$ classes.
    Images are first average-pooled and subsequently projected to a $5$-dimensional PCA representation before normalization to the interval $[-\pi,\pi]$. For the FashionMNIST o.o.d. dataset, the same preprocessing pipeline is applied.
    All remaining hyperparameters and architectural choices are identical to the MNIST setting.

    For each prediction, the volume of the singleton credal set was computed classically from the estimated singleton beliefs and plausibilities using an exact inclusion-exclusion formula for the volume of a simplex~\cite{lasserre2015volume}.
    
    The MNIST and CIFAR models achieve in-distribution test accuracies of 60.3\% and 34.3\%, respectively.
    The goal of the experiments is not to optimize predictive performance, but rather to evaluate the behavior of the proposed DST-based uncertainty measures in a proof-of-concept setting.
    \\

    \paragraph{Mass fusion from multiple classifiers.}
    To evaluate the effect of Dempster-Shafer mass fusion, we consider a multi-view classification setup based on the MNIST dataset.
    Each input image is partitioned into four disjoint patches of equal size, and a separate model is trained on each patch. For each patch, the input is reduced to match the number of qubits using principal component analysis (PCA) and subsequently rescaled to the interval $[-\pi,\pi]$. We use $n_{\mathrm{train}} = 300$ training samples and $n_{\mathrm{test}} = 300$ test samples.
    
    Each classifier is implemented as a parameterized quantum circuit with the same architecture as described in the previous paragraphs, using $5$ layers with AngleEmbedding for data encoding and nearest-neighbor entangling gates. 
    
    Training is performed using mini-batch stochastic optimization with batch size $16$ over $50$ epochs. We use the Adam optimizer with learning rate $0.1$ and set the loss parameter to $\lambda = 0.8$.
    The individual mass functions obtained from the trained classifiers are fused using Dempster-Shafer combination rules during evaluation.

    \subsection{Empty-set mass}
\begin{figure*}[t]
        \includegraphics[width=\textwidth]{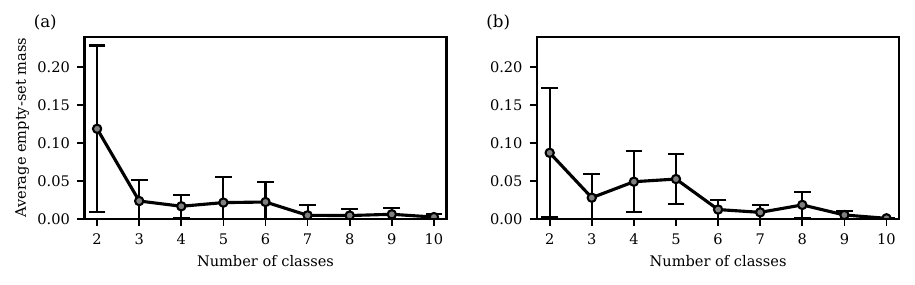}
        \caption{Average empty-set mass as a function of the number of classes.
        (a) MNIST and (b) CIFAR10. Error bars denote the standard deviation across test samples. Although the proposed encoding does not explicitly enforce the Dempster-Shafer constraint $m(\emptyset)=0$, the learned mass assigned to the empty set decreases with the number of classes and is negligible for the larger-scale settings considered in the main experiments.}
        \label{fig:emptysetmass}
    \end{figure*}

    In contrast to what is required by standard Dempster-Shafer theory, the encoding in Eq.~\eqref{eq:mass_from_rho} does not explicitly enforce $m(\emptyset)=0$, since the empty set corresponds to the all-zero computational basis state. 
    Figure~\ref{fig:emptysetmass} shows the average mass assigned to the empty set as a function of the number of classes.
    Although nonzero empty-set masses can occur, the learned mass assigned to the empty set decreases rapidly with problem size and becomes negligible in the regimes considered.

    The details of the experiment are as follows:
    We consider a multi-class classification task based on the MNIST (CIFAR10) dataset with $n$ classes, where $n$ is varied from 2 to 10.
    The input data is reduced to $n$ dimensions using principal component analysis (PCA) and subsequently rescaled to the interval $[-\pi, \pi]$ to match the range required for quantum feature encoding.
    We use $n_{\text{train}} = 300$ training samples and $n_{\text{test}} = 300$ test samples.

    The model is implemented as a PQC of $n$ qubits consisting of $5$ layers. 
    Classical input data is encoded using the AngleEmbedding routine provided by PennyLane. Each layer $k$ of the circuit consists of parameterized single-qubit rotations followed by entangling operations. Concretely, for each qubit $i$ and layer $k$, we apply
    \begin{align}
    \mathrm{RX}(\vartheta_{k i 1})\,\mathrm{RY}(\vartheta_{k i 2})\,\mathrm{RZ}(\vartheta_{k i 3}),
    \end{align}
    followed by controlled-NOT (CNOT) gates between neighboring qubits.    
    Training is performed using mini-batch stochastic optimization with batch size $16$ over $50$ epochs. 
    We use the Adam optimizer with learning rate $0.1$ to minimize the loss defined in Eq.~\eqref{eq:pl_sglloss}. The hyperparameter is set to $\lambda=0.8$.

\end{document}